\documentclass[aip,pop,graphicx]{revtex4-1}
\usepackage{graphicx}
\usepackage{float}
\makeatother

\begin{document}

\title{Pressure dependence of magnetron sputtering: 2D-RZ particle-in-cell and 1D fluid modeling}
\author{Joseph G. Theis}
\email{joseph.theis@colorado.edu}
\affiliation{Center for Integrated Plasma Studies, University of Colorado, Boulder, Colorado 80309, USA}
\author{Gregory R. Werner}
\affiliation{Center for Integrated Plasma Studies, University of Colorado, Boulder, Colorado 80309, USA}
\author{Thomas G. Jenkins}
\affiliation{Silvaco Inc., 5621 Arapahoe Avenue Suite A, Boulder, Colorado 80303, USA}
\author{Daniel Main}
\affiliation{Silvaco Inc., 5621 Arapahoe Avenue Suite A, Boulder, Colorado 80303, USA}
\author{John R. Cary}
\affiliation{Center for Integrated Plasma Studies, University of Colorado, Boulder, Colorado 80309, USA}
\affiliation{Silvaco Inc., 5621 Arapahoe Avenue Suite A, Boulder, Colorado 80303, USA}
\date{\today}

\begin{abstract}
We reproduce the consistently-seen experimental voltage versus pressure ($V$-$p$) dependence of DC magnetron sputtering (DCMS) with 2D-RZ particle-in-cell (PIC) simulation. 
Informed by PIC simulation, we develop a steady-state, 1D-axial fluid model of the sheath and presheath that also reproduces this $V$-$p$ dependence.
The $V$-$p$ dependence is the relationship between the steady-state voltage needed to maintain a constant discharge current and the neutral gas pressure.
$V$-$p$ dependence is fundamental to device performance, but has not previously been reproduced with simulation or satisfactorily explained. 
We find that the decrease in voltage with increasing pressure is not due to electron recapture at the cathode. 
Rather, the constant current dictates a constant global ionization rate, so the voltage decrease compensates for the increase in neutral gas density by lowering the energy of the plasma electrons, which decreases their ionization probability.
The PIC simulations also reveal that the presheath and bulk plasma are unaffected by the electron reflection coefficient at the cathode; the only effect of increasing reflection is a reduction in the sheath voltage and width.
In addition to the potential structure, we explore how pressure affects the plasma density, particle drifts, and particle energy distributions.
\end{abstract}

\maketitle

\section{Introduction} \label{sec:intro}

Magnetron sputtering is a form of plasma-based physical vapor deposition that is widely used to fabricate integrated circuits, photovoltaics, and other applications requiring metallic and compound coatings.\cite{brauer2010magnetron, gudmundsson2020physics, rossnagel2020magnetron}
Essentially, magnetron sputtering is a glow discharge formed between electrodes, wherein a magnetic field, applied perpendicular to the electric field, traps the electrons, allowing for operation at lower pressures and voltages.\cite{gill1965efficient}
The magnetized electrons ionize the neutral gas, producing unmagnetized ions that bombard the target material at the cathode, causing secondary electron emission and sputtering target atoms, which are then deposited on a substrate.
This $\bf E \times B$ discharge is similar to those in other devices, including Hall thrusters, magnetized capacitively-coupled plasmas, and Penning discharges.\cite{kaganovich2020physics}

Many types of magnetron sputtering devices exist with different electrode geometries and applied electric and magnetic fields.\cite{gudmundsson2020physics}
One of the most common configurations is the planar magnetron, in which a flat target (the cathode) has permanent magnets mounted behind it to create a region between the electrodes with a roughly perpendicular electric and magnetic field.
When this configuration is operated with a steady negative bias applied to the target, it is referred to as planar direct-current magnetron sputtering (DCMS).
Despite the widespread use of DCMS and other $\bf E \times B$ devices, much of our understanding remains empirical and many phenomena remain unexplained.\cite{gudmundsson2020physics,kaganovich2020physics}
This is due, in part, to the sensitivity of the discharge to external probes.\cite{bradley2001measurement, rossnagel1986langmuir,rossnagel2020magnetron}

Two of the most experimentally accessible and vital discharge parameters are the discharge voltage $V$ and the neutral gas pressure $p$.
Multiple studies have measured the discharge voltage as a function of pressure while maintaining a constant discharge current $I$, which we refer to as the $V$-$p$ dependence.\cite{thornton1978magnetron,westwood1983current,chang1986high,depla2006discharge}
While the magnitude of $V$ varies between studies due to different magnetic fields, currents, and materials, the variation in $V$ as a function of $p$ is very consistent. 
The voltage decreases monotonically with increasing pressure, decreasing rapidly until $\sim 3$ mTorr ($\approx 0.4$ Pa) and then decreasing more gradually.
This $V$-$p$ dependence is highly relevant to device operation, because pressure can affect the deposition rate, film quality, and input power.\cite{rossnagel1987pressure}
The $V$-$p$ dependence also provides a consistent experimental result against which models can be validated.

The leading explanation for the monotonic decline in voltage with increasing pressure has been based on electron recapture.\cite{thornton1978magnetron,buyle2003simplified,depla2006discharge,depla2009magnetron,ryabinkin2021structure}
When an ion strikes the cathode, an average of $\gamma$ electrons are emitted.
These cathode-sourced electrons (CSEs) are accelerated in the sheath, but can return to the cathode due to the generally radial magnetic field.
The electrons that return to the cathode have low energy (several eV), and are either absorbed or back-reflected; we define $C_{\rm r}$ as the probability of back-reflection.\cite{babout1977mirror,cimino2015detailed} 
Reflected electrons may similarly return to the cathode, where they may be again reflected or absorbed. 
Electrons that are eventually absorbed are called ``recaptured;'' we define the recapture probability as $f_{\rm r}$. 
Electrons that lose energy in an ionizing collision cannot return to the cathode, due to the sheath electric field.
Therefore, electrons that are recaptured contribute no net current, resulting in an effective ion-induced secondary electron emission yield $\gamma_{\rm eff} \equiv \gamma (1-f_{\rm r}) < \gamma$.\cite{lieberman2005principles, buyle2003simplified}

Recapture (hence $f_{\rm r}$) decreases with increasing pressure because electron-neutral collisions result in a net transport of electrons away from the cathode due to the axial electric field.
The Thornton equation\cite{thornton1978magnetron} is a particle balance model that assumes all ionization is performed by CSEs that gain energy in the sheath, yielding $V \propto \gamma_{\rm eff}^{-1} = (\gamma (1-f_{\rm r}))^{-1}$.
This framework has led past studies to claim that voltage decreases with increasing pressure because increasing pressure reduces recapture.\cite{thornton1978magnetron,buyle2003simplified,depla2006discharge,depla2009magnetron,ryabinkin2021structure}
We will show that this is not the primary explanation for the decrease in voltage with increasing pressure.

Since the introduction of the Thornton equation, many studies have found that electron heating in the presheath contributes substantially to the total ionization rate.\cite{buyle2005simplified, huo2013sheath, brenning2016role}
Studies often distinguish between the sheath heating experienced by the CSEs and the presheath heating experienced by all electrons, which has been variably referred to as Ohmic heating, Joule heating, or Hall heating.\cite{eremin2023electron}
In this work, we directly probe the influence of the sheath and presheath on ionization and study how these regions depend on pressure and cathode surface conditions.

Many approaches have been used to model DCMS, varying in complexity from global models to self-consistent 3D simulation.\cite{gudmundsson2020physics}
1D fluid models can offer the simplicity and interpretability of global models, while also producing accurate, quantitative results along the axial direction.
In these models, moments of the Boltzmann equation define a system of equations for the electrons and ions. 
Various closures and approximations can be applied to achieve analytical or numerical solutions.
This fluid approach has also been referred to as analytical\cite{lafleur2024analytical,gudmundsson2020physics} or moment analysis.\cite{surendra1993moment}
Such fluid models have the advantage of fast computation and can provide insights into the physics underlying the discharge.
Previous work has focused on modeling various aspects of the sheath, presheath, and bulk plasma, with different works adopting different models of electron cross-field transport and ionization.\cite{bradley1998study, bradley1997model, cramer1997analysis}
These studies have captured some qualitative trends, but have yielded only limited quantitative agreement with experiment and have not explained the $V$-$p$ dependence.

Particle-in-cell (PIC) simulations can self-consistently capture the spatial variations and non-Maxwellian particle distributions present in DCMS.\cite{birdsall1991particle}
Many studies have exploited the azimuthal symmetry of DCMS devices to model the discharge with 2D-RZ PIC simulation.\cite{kondo2001axisymmetrical,kolev2005influence,main2026modeling}
\citet{kolev2005influence} studied the dependence of DCMS on pressure and the cathode electron reflection coefficient $C_{\rm r}$ using 2D-RZ PIC simulation with a fixed external circuit.
Varying $C_{\rm r}$ directly affects recapture because reflection increases the probability that a CSE collides with the neutral gas and is transported away from the cathode.
\citet{ryabinkin2021structure} measured the $V$-$p$ dependence of DCMS with 2D-RZ PIC simulations. 
Their simulated $V$-$p$ curve was non-monotonic, exhibiting a maximum in $V$ at 1 mTorr, a minimum at 3 mTorr, and an increase in $V$ from 3 to 10 mTorr.
In contrast, experiments consistently report a monotonic decrease of $V$ with increasing $p$, a trend that has not previously been reproduced with self-consistent simulation.

In this work, we explore the pressure dependence and recapture dependence of DCMS using 2D-RZ PIC simulations.
The resulting $V$-$p$ dependence captures the monotonic decline and convexity of the experimental $V$-$p$ results, a first for PIC simulation.
By varying the recapture rate (via $C_{\rm r}$), while keeping pressure constant, we show that recapture is not responsible for this $V$-$p$ dependence.
To explain the $V$-$p$ dependence, we develop a 1D axial fluid model of the sheath and presheath regions.
The fluid model isolates a minimal set of physical effects to explain the basic presheath characteristics, and ultimately the V-p dependence in a way that is difficult with PIC simulation alone.
The fluid model shows that voltage decreases with increasing pressure because the total ionization rate in the presheath is roughly constant across pressures, so the ionization rate coefficient $k_{\rm iz}$ must compensate for the change in pressure. 
Since $k_{\rm iz}$ increases monotonically with electron energy, discharges at higher pressures can be sustained with lower electron energies, hence with lower voltages.
The fluid model also elucidates the relevant electron transport and energization mechanisms in this regime.
With boundary conditions roughly informed by PIC simulation, the fluid model accurately reproduces the DCMS $V$-$p$ dependence and the axial profiles of the fluid parameters.

In the next section, we describe the setup of our PIC simulations. 
In Sec.~\ref{sec:results}, we present the simulation results, highlighting the monotonic $V$-$p$ dependence and comparing with experiment. 
Section~\ref{sec:pressure} describes the sweep over pressures with constant $C_{\rm r}$; then Sec.~\ref{sec:refl} describes the sweep over electron reflection $C_{\rm r}$ with constant pressure.
Informed by the PIC results, we then develop a 1D-axial fluid model of the discharge in Sec.~\ref{sec:theory} and use it to explain the $V$-$p$ dependence. 
Finally, we summarize our work in Sec.~\ref{sec:conc}.

\section{PIC Simulation Setup} \label{sec:method}

\begin{figure}[H]
\centering
\frame{\includegraphics[scale=0.7]{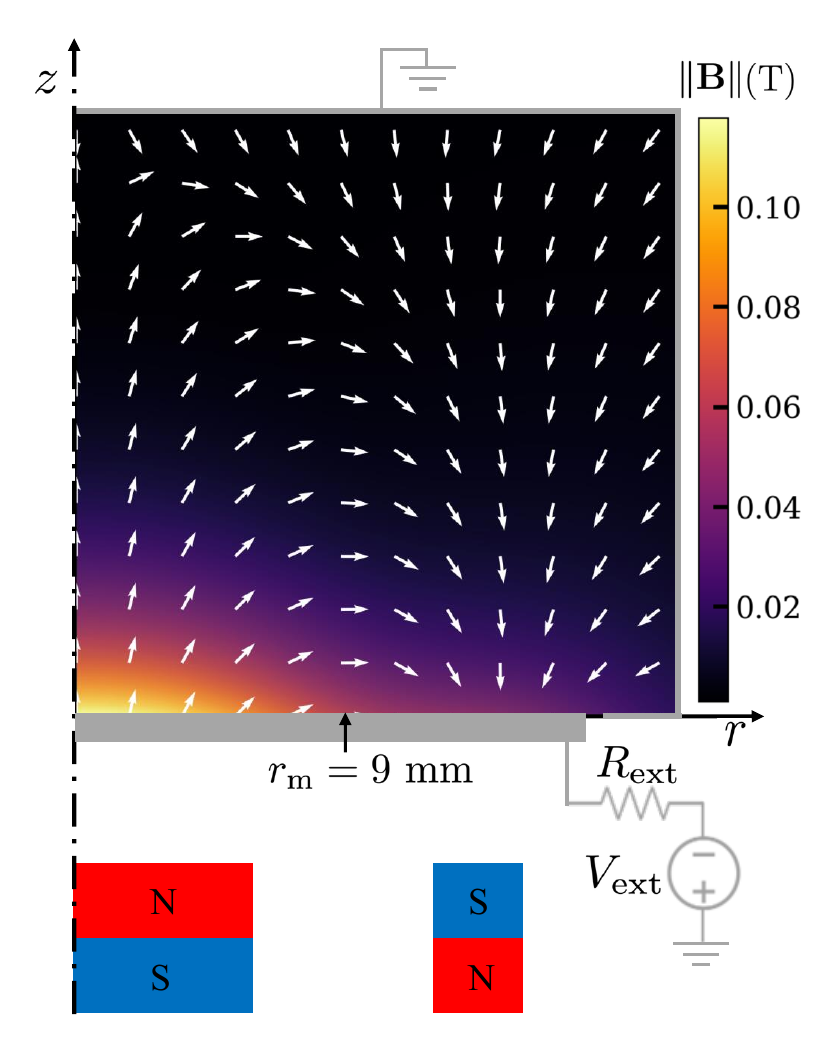}}
\caption{\label{fig:device} The 20x20 mm 2D-RZ cylindrical PIC domain displaying the magnetic field. The magnitude of the field is given by the color, while the arrows show direction. Outside of the PIC domain, the permanent magnets are drawn to scale. The cathode voltage is set according to the external circuit (shown in grey). All other walls are grounded. The black arrow at the cathode indicates the location of maximum ion flux $r_{\rm m}=9$ mm.}
\end{figure}

We simulated planar DCMS with the PIC code Vorpal\cite{nieter2004vorpal} (www.txcorp.com/vsim). 
Figure~\ref{fig:device} shows the 2D-RZ cylindrical PIC domain. The domain extended 20 mm radially and 20 mm axially, with the origin at the center of the cathode surface.
The cathode had a radius of 17 mm and was negatively biased by an external circuit with $V_{\rm ext}=-800$ V applied through a ballast resistance $R_{\rm ext}$, which was adjusted such that all simulations had a current of $I=100$ mA. 
The discharge voltage was, therefore, given by $V = -V_{\rm ext} - I R_{\rm ext}$, where we have defined the discharge voltage as positive, despite the cathode being negatively biased.
The other walls were grounded, and a one-cell discontinuous jump in voltage was imposed at the cathode–wall junction to represent the electrical separation between the biased cathode and a grounded ring.
The simulations were electrostatic.

\begin{figure}[H]
\centering
\frame{\includegraphics[scale=0.7]{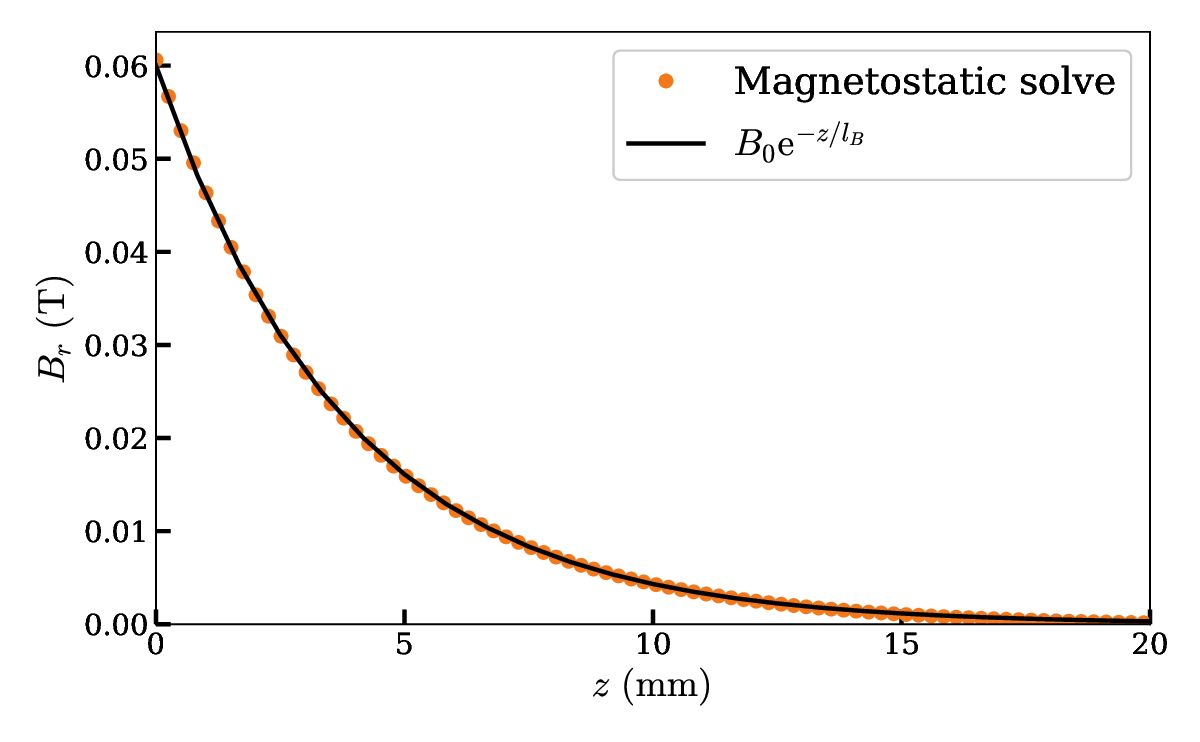}}
\caption{\label{fig:bLine} A lineout of the radial component of the magnetic field $B_r$ at $r=9$ mm. The orange dots show the result of the magnetostatic solve used in the PIC simulations. The solid black line shows the analytical form that was used in the fluid model (Sec.~\ref{sec:flu}), where $B_0=0.06$ T and $l_B=3.8$ mm.}
\end{figure}

All our PIC simulations used the same constant magnetic field profile, which we obtained from a numerical magnetostatic solve (also using Vorpal) on a larger domain that encompassed the two external permanent magnets.
Figure~\ref{fig:device} shows the position and polarity of the magnets and the resulting magnetic field in the PIC domain.
Both magnets had a remnant magnetization of 1.0 T and extended from -10 mm to -5 mm axially.
The central magnet was a cylinder with a radius of 6 mm.
The outer magnet was an annulus with an inner radius of 12 mm and an outer radius of 15 mm.
The magnetic flux from the annular magnet was greater than the flux from the cylindrical magnet, giving an unbalanced type II magnetic field structure.\cite{window1986charged} 
A lineout of the radial component of the magnetic field $B_r$ at $r=9$ mm is provided in Fig.~\ref{fig:bLine}, where $B_r$ decays nearly exponentially from its 0.06 T maximum at the cathode surface.

On average, the cathode emitted $\gamma = 0.07$ electrons per absorbed ion.\cite{phelps1999cold}
These CSEs were emitted with an energy of 4 eV directed axially, a negligible amount of energy compared to the sheath gain. 
Electrons that returned to the cathode were either absorbed (i.e. recaptured) or specularly reflected.
For the pressure sweep, the reflection coefficient was set to $C_{\rm r}=0.3$ (Refs. \onlinecite{babout1977mirror,cimino2015detailed}).
In Sec.~\ref{sec:refl}, we vary $C_{\rm r}$ to explore the dependence of the discharge on recapture.
All other boundaries simply absorbed particles.
We did not simulate the sputtered atoms.

The domain was filled uniformly with neutral argon gas at 300 K. 
The density of the neutral gas $n_{\rm g}$ was the parameter we changed to explore the pressure dependence of the discharge.
We included three kinds of electron-neutral collisions via a Monte Carlo Collision (MCC) algorithm\cite{birdsall1991particle}: elastic, ionization, and excitation.
For elastic collisions, the electrons were scattered according to the Vahedi–Surendra algorithm.\cite{vahedi1995monte}
For ionization collisions, the products were generated according to an algorithm developed by Kutasi and Donko.\cite{kutasi2000hybrid} 
For excitation collisions, the electrons were scattered isotropically with an energy reduced by the excitation threshold $\varepsilon_{\rm x}=11.5$ eV.
We used energy dependent cross-sections from the Hayashi and Phelps databases.\cite{hayashi_lxcat,phelps_lxcat}
For mTorr Ar pressures, the mean free path for electron-neutral collisions is long compared to the device size, but the magnetic field traps the electrons in $\bf E \times B$ trajectories, making these collisions relevant.
The ions are unmagnetized, so they travel directly to the cathode. This distance is much shorter than the ion-neutral collisional mean free path, so we neglected ion-neutral collisions.

We ran convergence studies to determine the appropriate macroparticle weight $w_{\rm p}$, timestep $\Delta t$, and cell size $\Delta z = \Delta r$. 
We used $V$ as the convergence metric and adopted resolutions such that $V$ was converged to within 5\% of its estimated asymptotic value. 
For all pressures, we used $w_{\rm p}=4 \times 10^4$, $\Delta t = 1.34$ ps, and $\Delta z= \Delta r=25.2$ $\rm \mu m$. 
The macroparticle weight resulted in $\sim 2 \times 10^7$ total macroparticles in steady state, with an average number of macro-electrons per cell of 17 globally and 170 in the presheath. 
The timestep corresponded to the cell-crossing time of an electron with energy $e V$, where $e$ denotes the elementary charge.
The cell size was approximately equal to half the minimum Debye length, resulting in 795 cells in $z$ and $r$.

We ran these simulations on the Perlmutter supercomputer at NERSC, utilizing between 512 and 1024 computational cores per simulation. 
In DCMS, the plasma density is highly nonuniform, which causes a large computational load imbalance when using regular domain decomposition for parallelization.
To speed up computation, we implemented a recursive coordinate bisection method to split the domain in a way that balances the computational load between cores.\cite{wolfheimer2006parallel}
The decomposition was dynamically updated as the discharge evolved to steady state.
Recursive coordinate bisection reduced the computation time by an order of magnitude from evenly-spaced subdomains.
    
\section{PIC Simulation Results} \label{sec:results}
Before we delve into the pressure and recapture dependence, we will outline the general features of the simulated discharges.
The discharges were initialized with steady-state results of the $p=2$ mTorr simulation, which itself was seeded by emitting a 100 mA electron current from the cathode for 0.2 ${\rm \mu s}$.
This reduced the transient adjustment period to a couple $\rm\mu s$ before the discharge current and voltage settled into steady state. 
The final steady-state discharge did not depend on the seeding technique, as long as a discharge was achieved.
At high pressures, there was a persistent small-amplitude ($\sim 4$ V) oscillation in the sheath voltage about a constant average, with a period of $\sim 1$ ${\rm\mu s}$.
To accurately represent the steady-state behavior, all values reported in this paper are time-averaged over 1 ${\rm\mu s}$.

\begin{figure}[H]
\centering
\frame{\includegraphics[scale=0.7]{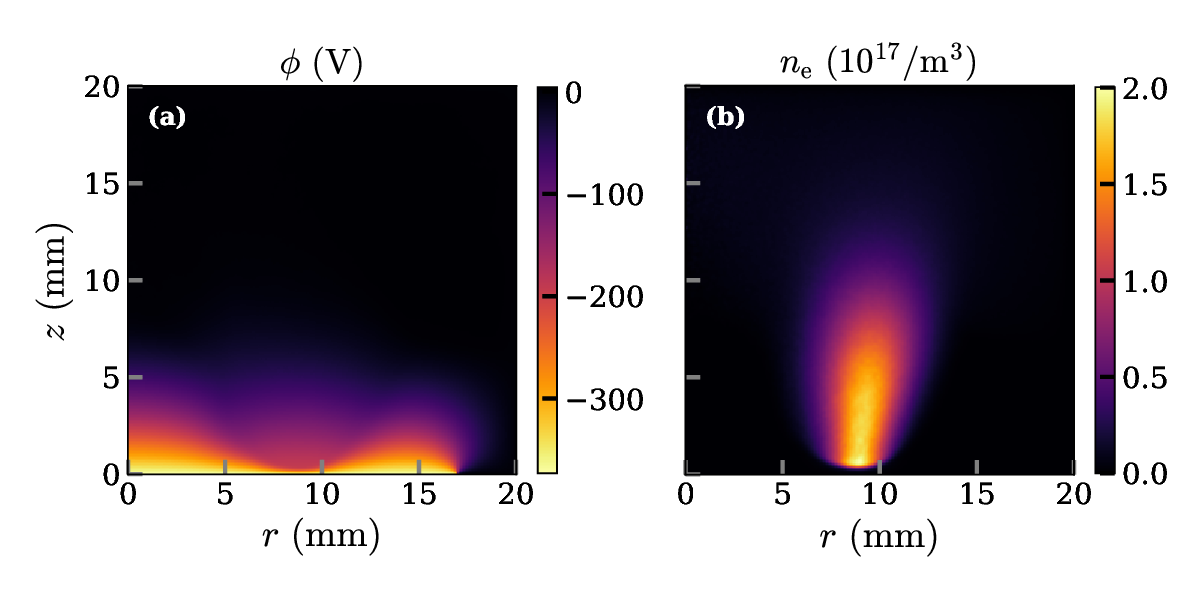}}
\caption{\label{fig:color} The 2D-RZ profiles of (a) the potential $\phi$ and (b) the electron density $n_{\rm e}$ for the PIC simulation with $p=2$ mTorr and $C_{\rm r}=0.3$.}
\end{figure}

\begin{figure}
\centering
\frame{\includegraphics[scale=0.7]{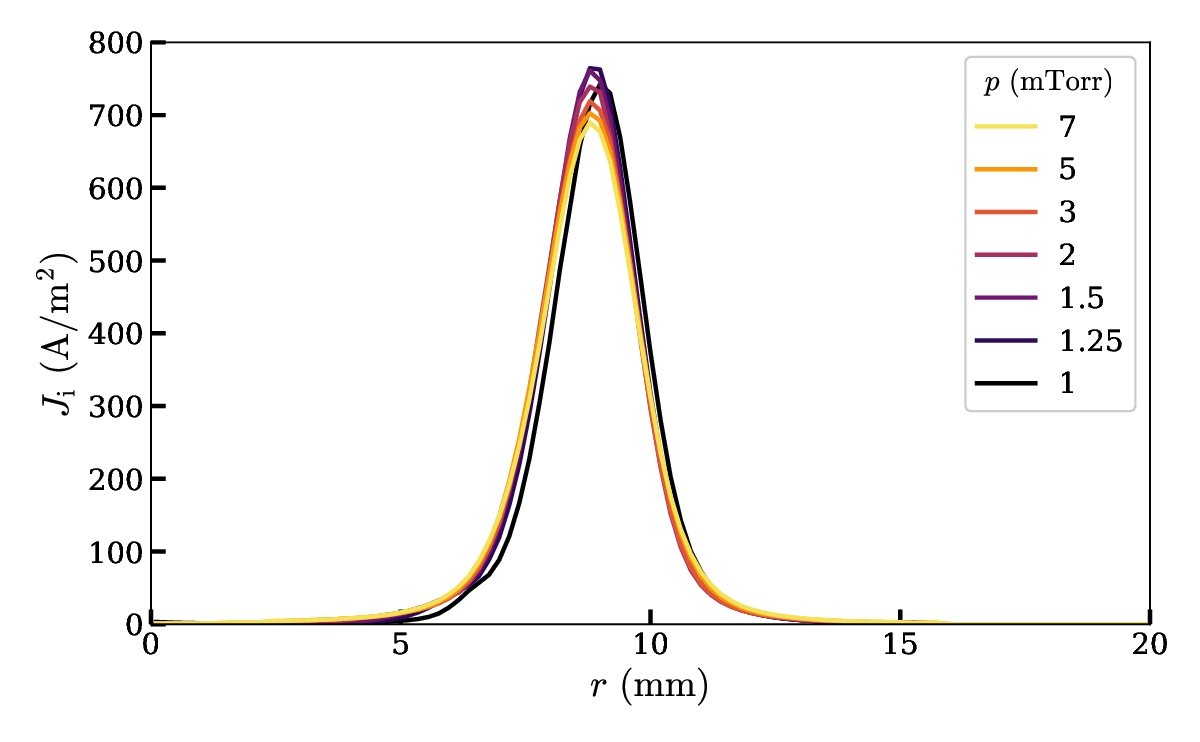}}
\caption{\label{fig:J} The ion current density along the cathode for the seven simulated pressures.}
\end{figure}

The steady-state potential $\phi$ is shown in Fig.~\ref{fig:color}(a) for the simulation with $p=2$ mTorr and $C_{\rm r}=0.3$.
The potential structure of DCMS consists of several distinct regions,\cite{bradley2001measurement,buyle2005simplified,lieberman2005principles} which we will describe from cathode to anode.
The sheath is the non-neutral region next to the cathode surface with the largest electric fields.
The presheath is the quasineutral region between the sheath and bulk plasma, with smaller, but significant electric fields.
The bulk plasma is the quasineutral region that extends from the presheath to the anode sheath, with a constant plasma potential $\sim 1$ V.
The anode sheath is the narrow non-neutral region where the plasma potential drops to ground.
In this study, we will focus primarily on the sheath and presheath, since, together, they constitute the total discharge voltage $V$.
The sheath width varies radially, narrowing where the magnetic field is tangential to the cathode surface.
The minimum sheath width $s \approx 0.5$ mm occurred roughly at the radius of maximum ion flux $r_{\rm m}=9$ mm (see Fig.~\ref{fig:J}).
As the sheath narrows, the width of the presheath grows, extending from the sheath edge to about $z=10$ mm.
A detailed, quantitative description of the sheath and presheath will be given in Sec.~\ref{sec:line}.

The corresponding electron density $n_{\rm e}$ is shown in Fig.~\ref{fig:color}(b).
The electrons are confined by the arching $\bf{B}$ field and axial $\bf{E}$ field. 
This results in the plasma density having an elongated peak in the presheath region that stays roughly centered above the $r_{\rm m}$. 
The only significant difference between $n_{\rm e}$ and the ion density $n_{\rm i}$ occurs in the sheath where $n_{\rm e}$ decays quickly towards the cathode, while $n_{\rm i}$ remains finite.
The plasma stays fairly confined radially, but does spread slightly as $z$ increases.

The ions generated in the discharge were accelerated by the electric field and bombarded the cathode along the characteristic racetrack sputtering profile.\cite{lieberman2005principles}
The bombarding ion flux peaked at $r_{\rm m}=9$ mm, where the magnetic field was parallel to the cathode surface.
The ion current density $J_{\rm i}$ at the cathode was roughly independent of pressure, as seen in Fig.~\ref{fig:J}.
Experimental measurements indicate that the pressure dependence of the racetrack width is sensitive to target erosion depth, target material, and magnetic field configuration. 
Depending on these parameters, the width has been observed to remain approximately constant, increase, or decrease with pressure.\cite{nakano2015growth,nakano2017transient,wendt1988dynamics}

\subsection{Neutral gas pressure dependence}\label{sec:pressure}
\begin{table*}[ht]
\caption{\label{tab:sims}Summary of simulation parameters and results. 
Above the line, we summarize simulations in which we vary $p$, while keeping $C_{\rm r}=0.3$. 
Below the line, we summarize simulations in which we vary $C_{\rm r}$, while keeping $p=2$ mTorr. 
The results of the pressure sweep are presented in Sec.~\ref{sec:pressure}. 
The $C_{\rm r}$ sweep will be discussed in Sec.~\ref{sec:refl}. 
The simulation with $C_{\rm r}=0.3$ and $p=2$ mTorr is listed twice.}
\begin{ruledtabular}
\begin{tabular}{ccccccccccc}
$p$ (mTorr) & $C_{\rm r}$ & $R_{\rm ext}$ ($\rm{k\Omega}$) & $V$ & $V_{\rm p}$ & $V_{\rm s}$ & $s$ (mm) & $n_{\rm p}$ ($\frac{10^{17}}{m^3}$) &  $\varepsilon_{\rm e,p}$ (eV) & $f_{\rm r}$\\
\hline 
1 & 0.3 & 3 & 507 & 342 & 165 & 0.56 & 1.1 & 57 & 0.97 \\
1.25 & 0.3 & 3.6 & 441 & 270 & 171 & 0.55 & 1.3 & 39 & 0.96 \\
1.5 & 0.3 & 3.9 & 409 & 231 & 178 & 0.54 & 1.4 & 31 & 0.95 \\
2 & 0.3 & 4.3 & 374 & 187 & 187 & 0.53 & 1.5 & 23 & 0.93 \\
3 & 0.3 & 4.6 & 344 & 141 & 203 & 0.54 & 1.7 & 17 & 0.89 \\
5 & 0.3 & 4.7 & 329 & 109 & 220 & 0.50 & 1.8 & 13 & 0.80 \\
7 & 0.3 & 4.78 & 322 & 89 & 233 & 0.50 & 2.0 & 11 & 0.74 \\
\hline
2 & 0 & 4.05 & 396 & 186 & 210 & 0.56 & 1.5 & 23 & 0.94 \\
2 & 0.3 & 4.3 & 374 & 187 & 187 & 0.53 & 1.5 & 23 & 0.93 \\
2 & 0.6 & 4.93 & 308 & 193 & 115 & 0.42 & 1.5 & 23 & 0.90 \\
2 & 0.9 & 5.45 & 255 & 200 & 55 & 0.29 & 1.5 & 23 & 0.82 \\
2 & 0.99 & 5.8 & 221 & 210 & 11 & 0.14 & 1.5 & 23 & 0.65
\end{tabular}
\end{ruledtabular}
\end{table*}

We simulated seven different pressures ranging from 1 to 7 mTorr, with $C_{\rm r}=0.3$. 
The input parameters and steady-state results for the pressure sweep are summarized in the first 7 rows of Table~\ref{tab:sims}.
We adjusted $R_{\rm ext}$ to maintain the constant 100 mA discharge current in all simulations. 
$V_{\rm p}$ and $V_{\rm s}$ are the presheath and sheath voltages, respectively.
Table~\ref{tab:sims} also lists the maximum plasma density in the presheath $n_{\rm p}$, maximum electron kinetic energy in the presheath $\varepsilon_{\rm e,p}$, and recapture probability $f_{\rm r}$.
We will address each column in the table over the course of Secs.~\ref{sec:results} and~\ref{sec:theory}.

\begin{figure}
\centering
\frame{\includegraphics[scale=0.7]{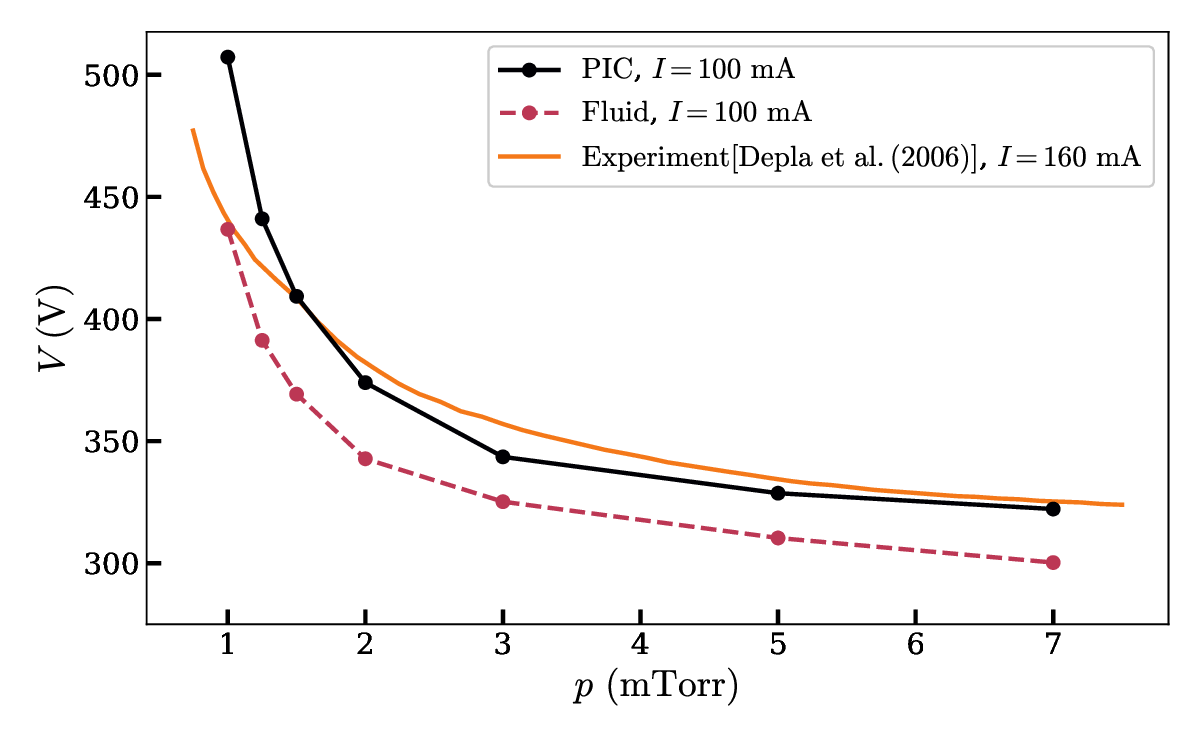}}
\caption{\label{fig:vp} $V$-$p$ curves obtained with our PIC simulation and 1D fluid model (presented in Sec.~\ref{sec:theory}) overlaid on experimental data~\cite{depla2006discharge} reprinted from Surface and Coatings Technology, Vol. 200, D. Depla, G. Buyle, J. Haemers, and R. De Gryse, “Discharge voltage measurements during magnetron sputtering,” pp. 4329–4338, \copyright~2006, with permission from Elsevier.}
\end{figure}

Importantly, in our PIC simulations, $V$ decreases monotonically with $p$.
As shown in Fig.~\ref{fig:vp}, $V$ decreases rapidly with pressure from 1 to $\sim 3$ mTorr and then decreases more gradually.
The magnitude of $V$ and the pressure dependence agree well with experimental results, including the $V$-$p$ measurements of \citet{depla2006discharge}, which we reprint in Fig.~\ref{fig:vp} for comparison.
Our simulated magnetron had a similar magnetic field and current density to the device in \citet{depla2006discharge}, but was $\sim 2/3$ the size and had $\sim 2/3$ the total discharge current, which sped up the simulations.
To our knowledge, this is the first PIC simulation to reproduce the $V$-$p$ dependence seen consistently in experiment.\cite{depla2006discharge,thornton1978magnetron,westwood1983current, chang1986high}

\subsubsection{Axial lineouts of potentials and densities}\label{sec:line}

\begin{figure}[H]
\centering
\frame{\includegraphics[scale=0.7]{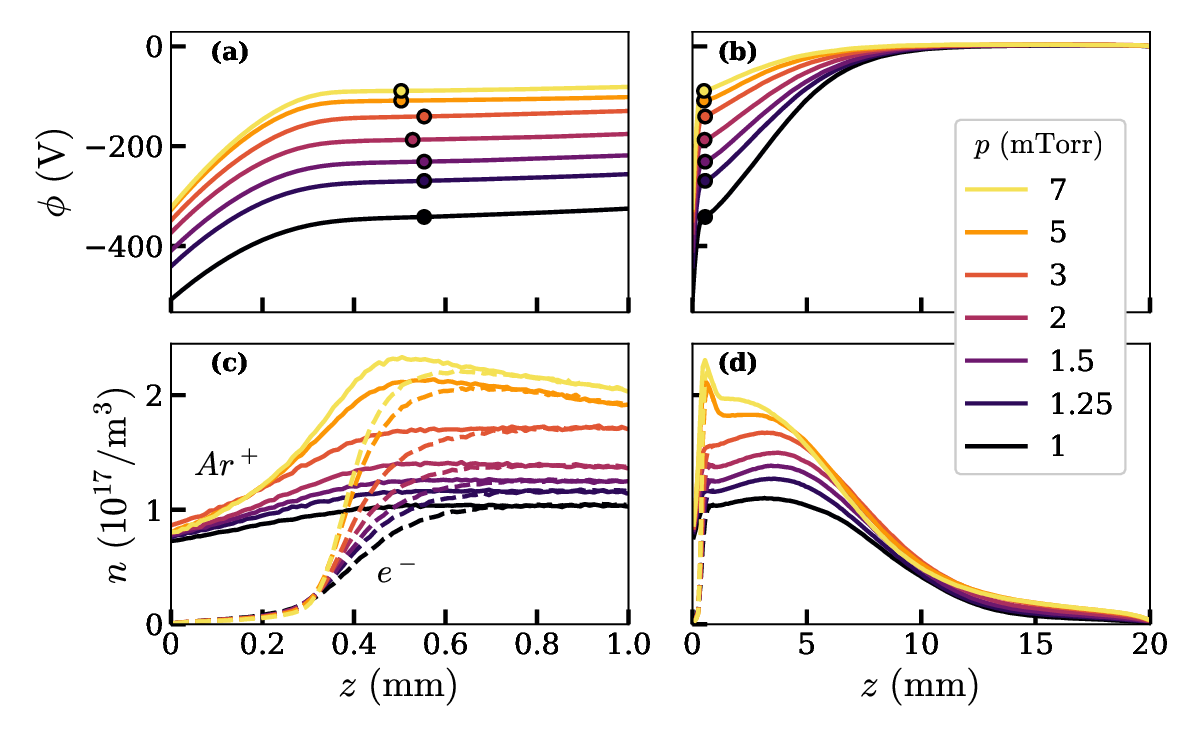}}
\caption{\label{fig:lineout} 
PIC lineouts of $\phi$, $n_{\rm i}$ (solid lines), and $n_{\rm e}$ (dashed lines) at $r=9$ mm for 7 pressures with $C_{\rm r}=0.3$. 
The right column shows the entire discharge gap, while the left column provides a detailed view of the first millimeter.
In panels (a) and (b), the sheath edge is marked by an $\circ$ filled with color corresponding to the pressure.
Outside of the narrow sheath, $n_{\rm i}=n_{\rm e}$.}
\end{figure}

Figure~\ref{fig:lineout} shows axial lineouts of $\phi$, $n_{\rm e}$, and $n_{\rm i}$ at $r=9$ mm.
The right column shows the entire discharge gap, including the sheath (0 to 0.5 mm), presheath (0.5 to 10 mm), and bulk (10 to 20 mm).
We define the boundary between presheath and bulk as the unique point where $\phi = 0$ V, which occurs around $z\sim10$ mm.
The left column provides a zoomed in view of the first millimeter, which contains the sheath.
Figure~\ref{fig:lineout}(c) clearly shows the non-neutral sheath transitioning to the quasineutral presheath at $z=s$, where $s \sim 0.5$ mm, roughly constant across all pressures.
The circles in the potential lineouts show the precise sheath edge, defined as the point where $n_{\rm e}=0.9n_{\rm i}$ (also see Table~\ref{tab:sims}).

Near the cathode, the electron density is much less than the ion density, because there are relatively few CSEs compared to ions, which carry nearly all the current at the cathode.
The electron density begins to increase rapidly around $z\sim0.3$ mm, leading to a narrow peak in electron and ion density around the sheath edge, which is more prominent for high pressures.
For perspective, the Larmor turning point of the CSEs immediately after emission (approximately twice their Larmor radius) occurs at  $z\sim 0.9$ mm.
The ion density increases more gradually from the cathode, affected only by acceleration in the sheath electric field and a negligible amount of ionization.
This positive charge density corresponds to the steep cathode sheath shown in Fig.~\ref{fig:lineout}(a), that gradually rolls over near the sheath edge.

Figure~\ref{fig:lineout}(d) shows that the discharge is quasineutral ($n_{\rm i} \approx n_{\rm e}$) outside of the narrow cathode sheath.
The presheath extends from the sheath edge to the position where $\phi\approx0$, which is roughly at $z\sim10$ mm for all pressures.
In the presheath, there is a broad peak in plasma density that decays toward the anode.
Figure~\ref{fig:lineout}(b) shows that the presheath has a significant electric field, that approaches zero near the presheath-bulk interface.
The bulk plasma potential is roughly constant ($\phi\sim1$ V), and the density is low compared to the presheath.

\begin{figure}
\centering
\frame{\includegraphics[scale=0.7]{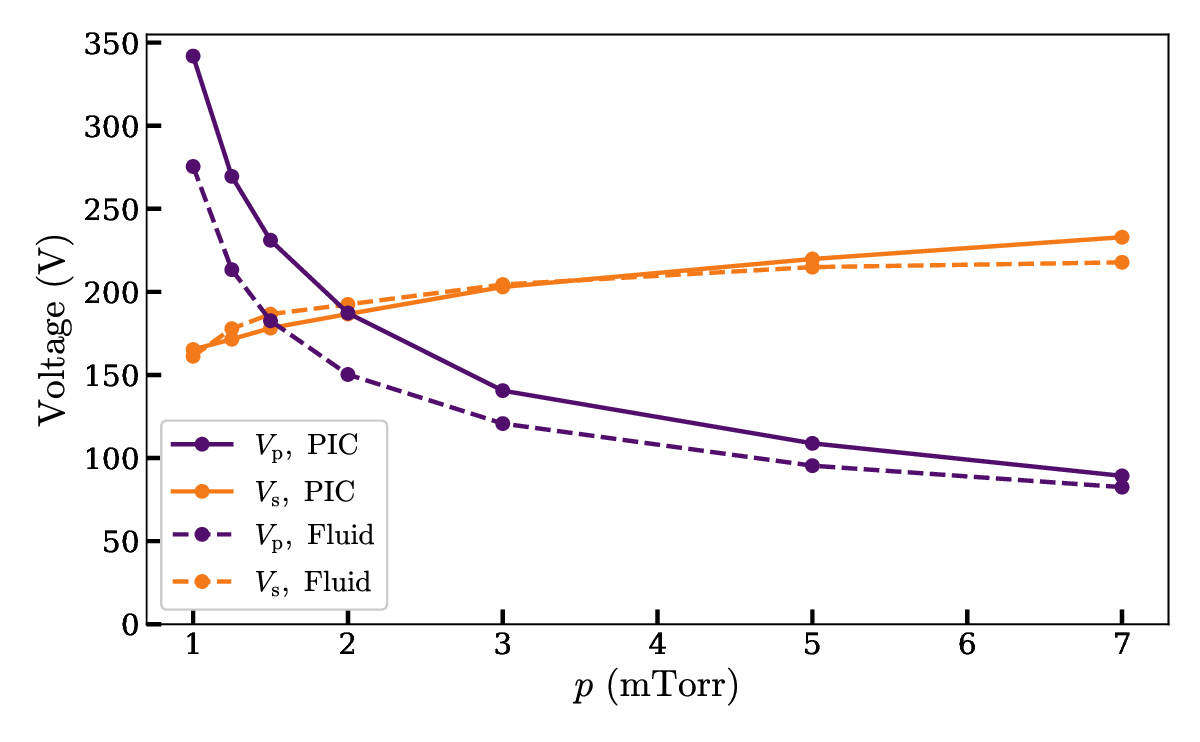}}
\caption{\label{fig:vsvp} The pressure dependence of the sheath and presheath voltages as calculated by PIC and the fluid model (presented in Sec.~\ref{sec:theory}).}
\end{figure}

The position of $s$ divides $V$ into its constituent parts, the sheath voltage $V_{\rm s}$ and presheath voltage $V_{\rm p}$.
The pressure dependence of $V_{\rm s}$ and $V_{\rm p}$ are shown in Fig.~\ref{fig:vsvp}.
The presheath voltage \emph{decreases} monotonically with increasing pressure, declining steeply at low pressures and gradually at high pressures.
The sheath voltage \emph{increases} monotonically with increasing pressure.
The variation in $V_{\rm p}$ with pressure is much greater than the variation of $V_{\rm s}$, hence the monotonic decline in $V$-$p$.
A similar dependence of the sheath and presheath voltages on pressure was observed by \citet{ryabinkin2021structure} in their PIC simulations, although their total $V$-$p$ dependence was non-monotonic.
The substantial presheath voltage indicates that electron heating in the presheath is important to sustaining the discharge, as has been noted in previous studies.\cite{buyle2005simplified,huo2013sheath,brenning2016role}

As pressure increases, the plasma density increases in the presheath and at the sheath edge.
This larger plasma density at the sheath edge leads to a larger ion and electron density in the sheath; however, the ion densities for all pressures are comparable at the cathode surface and the electron densities become comparable at $z \lesssim 0.3$ mm.
We will provide a theoretical explanation for these trends in Sec.~\ref{sec:theory}.

\subsubsection{Particle energy distributions}
\begin{figure}
\centering
\frame{\includegraphics[scale=0.7]{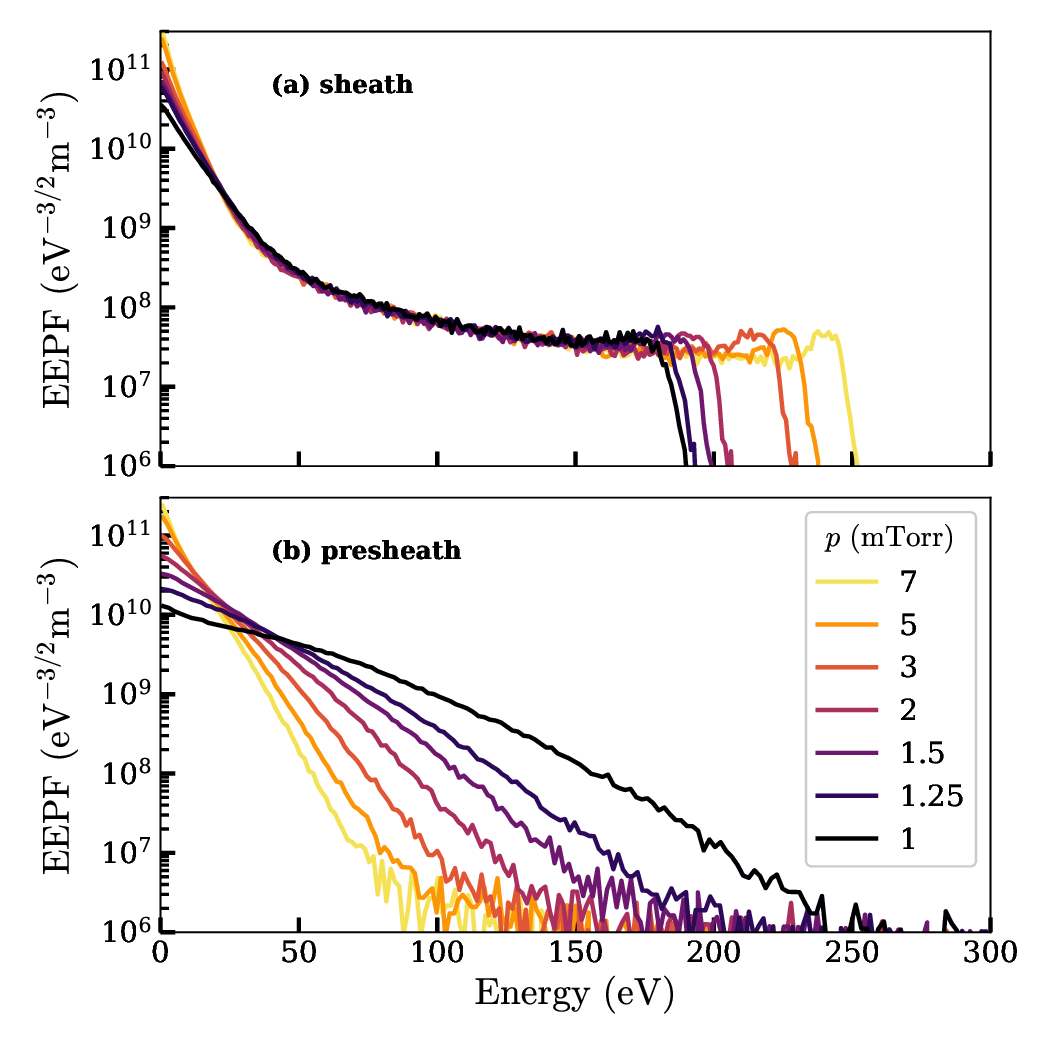}}
\caption{\label{fig:eepf} The electron energy probability functions for the 7 pressures in the volume extending from $r=$ 8 to 10 mm and (a) $z=$ 0 to 1 mm and (b) $z=$ 4 to 5 mm.}
\end{figure}

We measured the electron energy probability function\cite{lieberman2005principles} (EEPF) for the 7 simulated pressures in the sheath and presheath (see Fig.~\ref{fig:eepf}).
Given the scaling of the y-axis in Fig.~\ref{fig:eepf}, a Maxwellian distribution would yield a straight line with a slope equal to the negative, inverse temperature.
In the sheath [Fig.~\ref{fig:eepf}(a)], there is a roughly Maxwellian population of cold electrons and a high energy population that extends to the sheath energy gain $eV_{\rm s}$, where there is a bump in the distribution that corresponds to the uncollided CSEs.
The cold population consists primarily of ionization-sourced electrons, which are created outside the sheath, and therefore, have energies dictated by the smaller presheath electric field.
The temperature of the cold population decreases with pressure, from 9 eV at 1 mTorr to 5 eV at 7 mTorr due to the presheath electric field strength decreasing with increasing pressure. The inelastic collision rate also influences the temperature of this population, but this rate is roughly constant across pressures because the current is fixed, which fixes the total ionization rate (see Sec.~\ref{sec:flu}).

In the presheath, the EEPF is more uniform. 
Figure~\ref{fig:eepf}(b) shows the EEPF of a small, representative region near the center of the presheath.
At high pressures, the presheath distribution appears Maxwellian, while at low pressures the EEPF falls off faster than a Maxwellian because the azimuthal drift energy makes up a greater fraction of the electron kinetic energy at low pressures.
We see again that the average electron kinetic energy $\varepsilon_{\rm e}$ decreases with increasing pressure due to the decrease in the presheath electric field strength (exact values given in Table~\ref{tab:sims}).
With PIC simulation, \citet{ryabinkin2021structure} observed similar pressure dependence of the EEPF in the sheath and presheath.
Experimental measurements are limited in the sheath, but the presheath is typically found to be roughly Maxwellian.\cite{gudmundsson2020physics}

\begin{figure}
\centering
\frame{\includegraphics[scale=0.7]{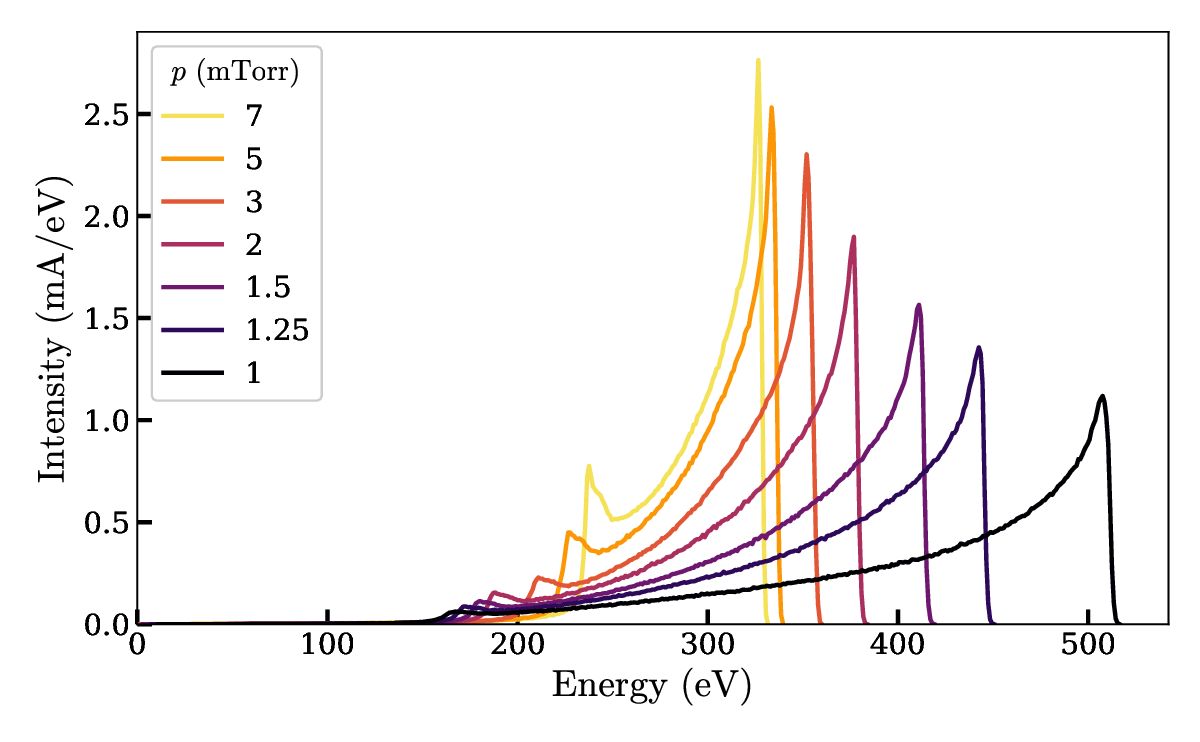}}
\caption{\label{fig:iedf} The energy distribution of ions absorbed at the cathode for the seven pressures.}
\end{figure}

\newpage
The energy distributions of the ions bombarding the cathode are provided in Fig.~\ref{fig:iedf}.
Since ions carry nearly all the current at the cathode, each distribution integrates to $\sim100$ mA.
The ions, being collisionless and unmagnetized, simply get accelerated by the electric field from their place of origin to the cathode. Thus, their bombarding energy spectra show the potential (relative to the cathode) where the ions were created.
Little ionization occurs in the sheath, where the CSEs are rapidly accelerated by the sheath electric field.
Near their Larmor turning point, the CSEs begin ionizing the neutral gas, causing a small peak in the spectra near $eV_{\rm s}$.
Following this small peak, there is a roughly exponential cascade of ionization through the presheath that leads to a sharp peak in the spectra at the full discharge voltage $eV$.
This results in a striking pressure dependence of the bombarding ion spectra, where the energy of the small peak increases with pressure due to increasing $V_{\rm s}$ and the energy of the large peak decreases with increasing pressure due to the decreasing $V$ (cf. Fig.~\ref{fig:vsvp}).
This result shows that the potential bounds of the presheath dictate the bombarding ion energy, with higher pressures producing a more narrow range of energies with a lower average.
This also demonstrates how pressure can affect figures of merit in DCMS because the energy of the bombarding ions influences the sputter yield and deposition rate of the device.\cite{yamamura1996energy,gudmundsson2020physics}
If we were to include ion collisions, the peaks would be slightly dulled, but the general structure of these spectra would be preserved.

\subsection{Electron reflection coefficient sweep}\label{sec:refl}
Previously, the decline of voltage with pressure has been attributed to the reduction of recapture with increasing pressure.\cite{thornton1978magnetron,buyle2003simplified,depla2006discharge,depla2009magnetron,ryabinkin2021structure}
We see in Table~\ref{tab:sims}, that $f_{\rm r}$ does in fact decrease with increasing pressure.
But does this variation in $f_{\rm r}$ account for the large decrease in $V_{\rm p}$ that we observed in the pressure sweep?
To disentangle the effect of recapture on the discharge from that of pressure, we varied the electron reflection coefficient $C_{\rm r}$ at the cathode with a constant $p=2$ mTorr.
$C_{\rm r}$ directly affects $f_{\rm r}$ because each reflection increases the probability that a CSE collides with the neutral gas and is transported away from the cathode.
Furthermore, it is instructive to measure the sensitivity of the discharge to $C_{\rm r}$ because the electron reflection coefficient is sensitive to the cathode surface condition and is difficult to measure experimentally.\cite{babout1977mirror,cimino2015detailed}

\begin{figure}
\centering
\frame{\includegraphics[scale=0.7]{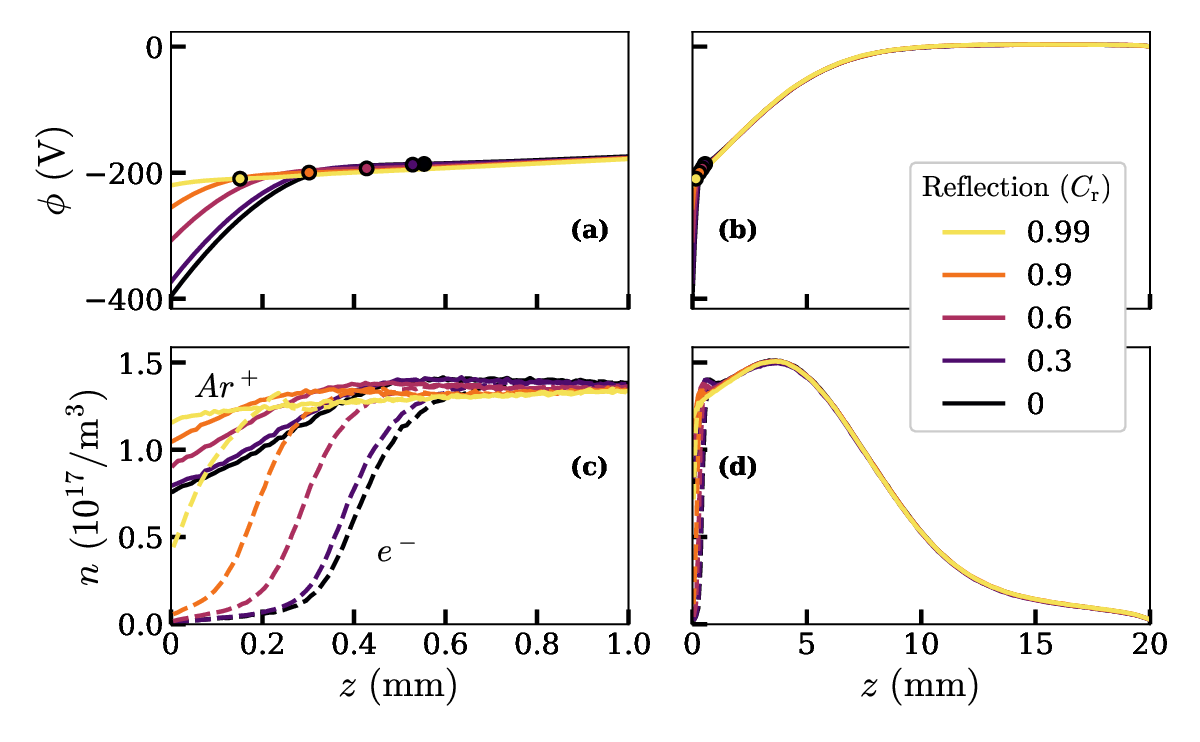}}
\caption{\label{fig:lineoutRefl} 
PIC lineouts of $\phi$, $n_{\rm i}$ (solid lines), and $n_{\rm e}$ (dashed lines) at $r=9$ mm for five different reflection coefficients with $p=2$ mTorr.
The right column shows the entire discharge gap, while the left column provides a detailed view of the first millimeter.
In panels (a) and (b), the sheath edge is marked by an $\circ$ filled with the color corresponding to the reflection coefficient.
Outside of the narrow sheath, $n_{\rm i}=n_{\rm e}$.}
\end{figure}

For $p=2$ mTorr, we simulated 5 different values of the reflection coefficient, given in Table~\ref{tab:sims}.
We determined $f_{\rm r}$ by measuring the ratio of the absorbed electron current to the emitted electron current at the cathode.\cite{ryabinkin2021structure}
As expected, $f_{\rm r}$ varied inversely with $C_{\rm r}$.
For $C_{\rm r}=0$, only 6\% of electrons escaped recapture. 
As $C_{\rm r}$ increased to 0.99, the escaped fraction increased to 35\% (setting $C_{\rm r}=1$ led to an unstable increase in $n_{\rm e}$ at the cathode surface, so we do not show this result).
Figure~\ref{fig:lineoutRefl} shows axial lineouts of $\phi$, $n_{\rm e}$, and $n_{\rm i}$ at $r=9$ mm for the 5 values of $C_{\rm r}$.
Increasing $C_{\rm r}$ caused the sheath width and sheath voltage to decrease.
This is due to increased reflection increasing the electron density at the cathode, which shrinks the non-neutral region of the discharge.
\citet{kolev2005influence} observed a similar increase in electron density at the cathode with increasing $C_{\rm r}$ in their PIC simulations of DCMS.
We see that the sheath nearly disappears for $C_{\rm r}=0.99$.

Although varying $C_{\rm r}$ strongly affected the recapture rate as anticipated, we found that it remarkably did not affect the plasma outside of the sheath.
This is evidenced by the nearly identical potential and density lineouts for $z>1$ mm in Fig.~\ref{fig:lineoutRefl}(b) and~\ref{fig:lineoutRefl}(d).
This result has important consequences for modeling the discharge, namely that the plasma parameters in the presheath and bulk can be robustly determined even without accurate boundary conditions for the cathode surface.

This reflection sweep taken together with the pressure sweep in Sec.~\ref{sec:pressure} indicates that the decrease in voltage with increasing pressure is not due to recapture as past studies proposed.\cite{thornton1978magnetron,buyle2003simplified,depla2006discharge,depla2009magnetron,ryabinkin2021structure}
Figure~\ref{fig:vsvp} shows that the monotonic decline in voltage with increasing pressure is a reflection of the decreasing presheath voltage.
Figure~\ref{fig:lineoutRefl} shows that the presheath is insensitive to large variations in $C_{\rm r}$ and recapture.
Therefore, it is unlikely that the decrease in $V$ with increasing pressure is due to the decrease in recapture with increasing pressure.
Furthermore, while reducing recapture via $C_{\rm r}$ decreases the sheath voltage, reducing recapture by increasing pressure (Sec. \ref{sec:pressure}) has the opposite effect on sheath voltage. 
This shows that the sheath is more strongly affected by the pressure dependence of the presheath than by the pressure dependence of recapture.
Studies of rf magnetron sputtering have shown that CSEs energized by the sheath play an \emph{auxiliary} role to the dominant presheath energization mechanisms.\cite{eremin2023electron,zheng2021electron} 
This appears to be the case for the DCMS discharge modeled here.
In the next section, we develop an explanation of $V$-$p$ that focuses on the influence of pressure on particle balance in the presheath.

\section{1D-axial fluid model}\label{sec:theory}
\subsection{1D-axial fluid presheath model}\label{sec:flu}
To understand the $V$-$p$ dependence observed in our PIC results and in previous experiments, we develop a steady-state, 1D-axial fluid model of the sheath and presheath regions.
We begin with the presheath, which we will boundary match to the sheath model at $z=s$ in the following subsection.
We assume the plasma is quasineutral ($n_{\rm e} = n_{\rm i} \equiv n$) and uniform in the radial and azimuthal directions in the cross-sectional area $A(z)$.
We take $\mathbf{E}=E\hat{z}$ and $\mathbf{B} = B \hat{r}$.

For electrons, we consider ionization, excitation, and elastic collisions with rate coefficients $k_{\rm iz}$, $k_{\rm x}$, and $k_{\rm e}$, respectively.
We assume Maxwellian electron distributions $f_{\rm M}$ with temperatures given by $2\varepsilon_{\rm e}/3$, so the rate coefficient for some collision $\alpha$ is 
\begin{equation}\label{eq:rate}
k_\alpha(\varepsilon_{\rm e})=\int_0^\infty f_{\rm M}(\varepsilon;\varepsilon_{\rm e})\sigma_\alpha(\varepsilon) \sqrt{\frac{2\varepsilon}{m}}d\varepsilon,
\end{equation}
where $\sigma_\alpha(\varepsilon)$ is the cross-section for collision $\alpha$, and $m$ is the electron mass.
We use the same cross-section data as the PIC simulations.

The ion current $I_{\rm i}=Aneu_{\rm i}$ and electron current $I_{\rm e}=-Aneu_{\rm e}$ are related to the total discharge current via
\begin{equation}\label{eq:Id}
I = I_{\rm i} + I_{\rm e} = Ane(u_{\rm i}-u_{\rm e}),
\end{equation}
where $u_{\rm i}$ and $u_{\rm e}$ are the axial ion and electron drifts, respectively.
We take $\hat{z}$ to point from the cathode to the anode as we did in the PIC simulations, yielding the following sign conventions: $I<0$,  $I_{\rm i}<0$, $I_{\rm e}<0$, $u_{\rm i}<0$, $u_{\rm e}>0$, and $E<0$.
The steady-state continuity equations including ionization as a source term are 
\begin{equation}\label{eq:cont}
\frac{dI_{\rm i}}{dz} = -\frac{dI_{\rm e}}{dz} = A e n n_{\rm g} k_{\rm iz}.
\end{equation}
The steady-state, axial, cold, unmagnetized ion momentum equation is
\begin{equation}\label{eq:ionp}
\frac{d}{dz}\left( I_{\rm i} u_{\rm i} \right) = Ae^2nE/M,
\end{equation}
where $M$ is the ion mass. 
We have assumed in Eq. \ref{eq:ionp} that ions are born with a velocity that is negligible compared to the drift $u_{\rm i}$.
We describe the axial electron transport classically, using Pedersen conductivity,\cite{pedersen1927propagation,lieberman2005principles} which balances electromagnetic and collisional drag forces in the steady-state momentum equation.
This yields an axial drift, here simplified to the case when the total collision frequency $n_{\rm g}k_{\rm c} \equiv n_{\rm g}(k_{\rm iz}+k_{\rm x}+k_{\rm e})$ is small compared with the cyclotron frequency $\omega_{\rm c}$.
\begin{equation}\label{eq:E}
u_{\rm e} = -\frac{en_{\rm g}k_{\rm c}}{m\omega_{\rm c}^2}E.
\end{equation}
For the steady-state electron energy equation,\cite{gurnett2005introduction} we approximate the heat flux as $Q=nu_{\rm e}\varepsilon_{\rm e}$, yielding
\begin{equation}\label{eq:qe}
\frac{d}{dz}(I_{\rm e} \varepsilon_{\rm e}) = -e I_{\rm e} E + Aenn_{\rm g} (k_{\rm iz} \varepsilon_{\rm iz} + k_{\rm x} \varepsilon_{\rm x}),
\end{equation}
where $\varepsilon_{\rm iz}$ and $\varepsilon_{\rm x}$ are the ionization and excitation energies, respectively.
The first term on the right-hand side is Joule heating and the second term represents the energy loss due to inelastic collisions.
We expand the derivatives and manipulate the above expressions to yield:
\begin{equation}\label{eq:ui}
\frac{du_{\rm i}}{dz} = \frac{eE}{Mu_{\rm i}} - n_{\rm g} k_{\rm iz}
\end{equation}
\begin{equation}\label{eq:ue}
\frac{du_{\rm e}}{dz} = n_{\rm g} k_{\rm iz} - \frac{u_{\rm e}}{u_{\rm i}}\left( 2 n_{\rm g} k_{\rm iz} -\frac{eE}{Mu_{\rm i}} \right)
\end{equation}
\begin{equation}\label{eq:ee}
\frac{d\varepsilon_{\rm e}}{dz} = -eE - \frac{n_{\rm g} k_{\rm iz}}{u_{\rm e}} (\varepsilon_{\rm e} +\varepsilon_{\rm iz} + k_{\rm x} \varepsilon_{\rm x}/k_{\rm iz}).
\end{equation}

The system is fully determined by the three nonlinear ODEs given in Eqs. (\ref{eq:ui}), (\ref{eq:ue}), and (\ref{eq:ee}) and the constraints given in Eqs.  (\ref{eq:rate}), (\ref{eq:Id}), and (\ref{eq:E}).
To approximate our specific discharge, we take the cross-sectional area to be $A(z) = 2 \pi r_{\rm m} w(1+z/l_A)$, where $w=3.4$ mm is the approximate full width at 25\% maximum of the peak in $J_{\rm i}(z=0)$ (see Fig.~\ref{fig:J}), and $l_A=10$ mm accounts for the radial spreading with increased $z$ observed in our PIC simulations.
We also take $\mathbf{B} = B_0 \mathrm{e}^{-z/l_B}\hat{r}$, with $B_0=0.06$ T and $l_B = 3.8$ mm (cf. Fig.~\ref{fig:bLine}).

We integrate this system numerically from the presheath-bulk interface at $z=10$ mm to the sheath edge at $z=0.52$ mm.
In the PIC simulations, the values of $u_{\rm e}$ and $u_{\rm i}$ at $z=10$ mm are insensitive to pressure. 
Thus, in the fluid model we set the boundary conditions at $z=10$ mm to $u_{\rm i,0}=-2.5\times10^{3}$ m/s and $u_{\rm e,0}=2.5\times10^4$ m/s for all pressures.
Integration is highly sensitive to the third boundary condition $\varepsilon_{\rm e,0}$ (the electron kinetic energy at $z=10$ mm).
Above some critical value of $\varepsilon_{\rm e,0}$, $\varepsilon_{\rm e}$ diverges to unphysically large values during integration.
Below the critical value of $\varepsilon_{\rm e,0}$, $\varepsilon_{\rm e}$ goes to zero.
The solutions remain stable across the entire domain only when the critical value of $\varepsilon_{\rm e,0}$ is found with high precision ($\sim 10^{-12}$ eV). 
We find this critical $\varepsilon_{\rm e,0}$ using a bisection search method, where we iteratively narrow the bounds on $\varepsilon_{\rm e,0}$ based on whether $\varepsilon_{\rm e}$ diverges to infinity or zero during integration, until we find a value that permits integration to the sheath edge.
Essentially, this precise value of $\varepsilon_{\rm e,0}$ suppresses numerical instabilities of the ODE system,\cite{press2007numerical} and yields the physical solution.
Thus, the extreme sensitivity to $\varepsilon_{\rm e,0}$ is a numerical feature of integrating this ODE system, not a feature of the physical or PIC discharge where stability is maintained over much greater fluctuations in plasma parameters.

\begin{figure}
\centering
\frame{\includegraphics[scale=0.7]{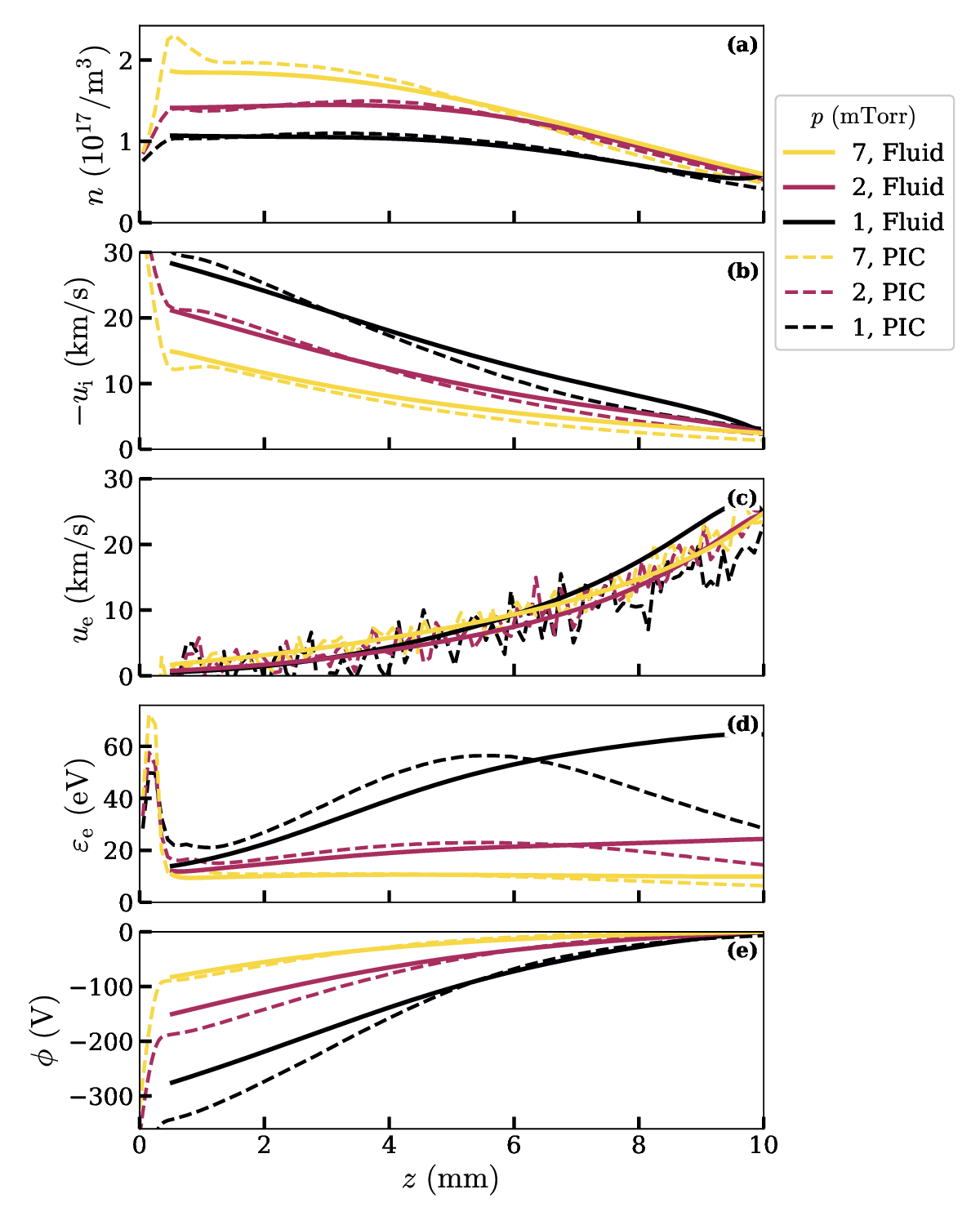}}
\caption{\label{fig:fluPre} Solutions of the 1D fluid presheath model for 1, 2, and 7 mTorr (solid lines), and the corresponding PIC lineouts at $r=9$ mm (dashed lines).}
\end{figure}

Solutions of the presheath fluid model are shown along with PIC lineouts in Fig.~\ref{fig:fluPre} for the pressures 1, 2, and 7 mTorr. 
The plasma densities produced by the fluid model agree with the PIC results to within 15\% [Fig.~\ref{fig:fluPre}(a)].
Some features at the sheath edge, like the sharp density peak for 7 mTorr, are due to CSEs, which constitute a separate non-Maxwellian population [see Fig. \ref{fig:eepf}(a)] that is not included in our presheath fluid model.
The strong pressure dependence of the ion drifts and weak pressure dependence of the electron drifts are accurately captured with the fluid model, with better agreement in the low-$z$ half of the domain [Figs.~\ref{fig:fluPre}(b) and~\ref{fig:fluPre}(c)].
The fluid model captures the pressure-dependent Joule heating of electrons, predicting the maximum energy to within 10\%, but the electron energy is overestimated near the bulk [Fig.~\ref{fig:fluPre}(d)].
In general, the model accurately predicts the fluid parameters of the presheath, with agreement improving as the sheath is approached. 
The collision frequencies are well approximated by the assumption of Maxwellian electron distributions.
The error near the presheath-bulk interface is likely due to the model neglecting the electrons' diamagnetic, grad-B, and curvature drifts, which increase as the Larmor radii grow with decreasing $B_r(z)$.

The potentials from the fluid model generally underestimate the PIC potentials, but still agree to within 20\% [Fig.~\ref{fig:fluPre}(e)].
The potential at $z=s$ gives the value of $V_{\rm p}$.
The pressure dependence of $V_{\rm p}$ for the fluid model is shown along with the PIC values in Fig.~\ref{fig:vsvp}. 
The fluid model predicts the value of $V_{\rm p}$ to within 20\% and captures the monotonic decline and convexity.

The fluid model helps to elucidate the constant-current pressure dependence of $V_{\rm p}$. 
In steady state, the total number of ions created per second (the global ionization rate) precisely equals the rate of ions hitting the boundaries, primarily the cathode.
The low ion-induced secondary electron yield at the cathode and the high recapture rate means that the discharge current is composed nearly entirely of ions at the cathode.
The small amount of ionization that occurs in the sheath means that the current at the sheath edge is also dominated by ions. 
Few ions are created in the bulk and there is no significant bulk electric field to drive an ion current, so the current at the presheath-bulk interface is dominated by electrons accelerated in the presheath.
This means that the total ionization rate in the presheath $S_{\rm iz,p}$ [the integral of Eq. (\ref{eq:cont})] will be roughly equal to $I$ for all pressures.

$S_{\rm iz,p}$ is proportional to $n_{\rm g}$, $n$, and $k_{\rm iz}$, which increases monotonically with $\varepsilon_{\rm e}$.
As $n_{\rm g}$ increases between simulations, $nk_{\rm iz}$ must compensate to maintain $S_{\rm iz,p} \approx I$.
This is accomplished by decreasing $V_{\rm p}$. 
Decreasing $V_{\rm p}$ decreases $\varepsilon_{\rm e}$ (as confirmed by PIC and the fluid model), which reduces $k_{\rm iz}$, especially when $\varepsilon_{\rm e} \lesssim \varepsilon_{\rm iz}$.
Decreasing $V_{\rm p}$ also decreases $u_{\rm i} \sim \sqrt{V_{\rm p}}$, which means that $n$ must increase weakly to maintain the roughly constant ion current at the sheath edge.
The decrease in $k_{\rm iz}$ dominates the increase in $n$, so $S_{\rm iz,p}$ can be kept constant as $n_{\rm g}$ increases by decreasing $V_{\rm p}$.

The convexity of $V_{\rm p}(p)$ is primarily due to the roughly Arrhenius form of $k_{\rm iz}(\varepsilon_{\rm e})$. 
Below the ionization energy, $k_{\rm iz}$ decays rapidly as $\varepsilon_{\rm e}$ decreases, becoming negligible around $\varepsilon_{\rm e} \lesssim 6$ eV. 
Above the ionization energy, $k_{\rm iz}$ grows sublinearly with increasing $\varepsilon_{\rm e}$.
As we argued above, higher pressures lead to lower electron energies (as confirmed in PIC and 1D fluid simulation [Fig.~\ref{fig:fluPre}(d)]).
Therefore, at high pressures, where electron energies are less than the ionization energy, a small amount of electron energization leads to a large relative change in $k_{\rm iz}$; while at low pressures, where electron energies are greater, more energization (and thus voltage) is required to effect the same relative change in $k_{\rm iz}$. 

\subsection{1D-axial fluid sheath model}
To complete our fluid model of the discharge, we apply the output of the quasineutral presheath model as input to a non-neutral sheath model.
The sheath model extends from the cathode surface to the point of quasineutrality at the sheath edge $(n_{\rm e}(s) = n_{\rm i}(s) \equiv  n_s)$. 
Based on our PIC results, the ionization rate in the sheath is small compared with the ion number current, so we treat the ion current as uniform within the sheath. 
Energy conservation for a cold ion beam gives
\begin{equation}\label{eq:ni}
n_{\rm i} = n_s \left(1 - \frac{2e(\phi+V_{\rm p})}{Mu_{{\rm i},s}^2}\right)^{-1/2}.
\end{equation}
We describe the electron density with the Boltzmann relation, assuming constant average energy $\varepsilon_{{\rm e},s}$:
\begin{equation}\label{eq:ne}
n_{\rm e} = n_s {\rm e} ^ {3e(\phi+V_{\rm p})/2 \varepsilon_{{\rm e},s}}.
\end{equation}
Plugging these expressions into Poisson's equation yields the second-order nonlinear differential equation\cite{lieberman2005principles}
\begin{equation}\label{eq:sheath}
\frac{d^2\phi}{dz^2} = \frac{e n_s}{\varepsilon_0} \left[ {\rm e}^{3e(\phi+V_{\rm p})/2 \varepsilon_{{\rm e},s}} - \left( 1 - \frac{2e(\phi+V_{\rm p})}{Mu_{{\rm i},s}^2} \right)^{-1/2}\right].
\end{equation}

\begin{figure}[H]
\centering
\frame{\includegraphics[scale=0.7]{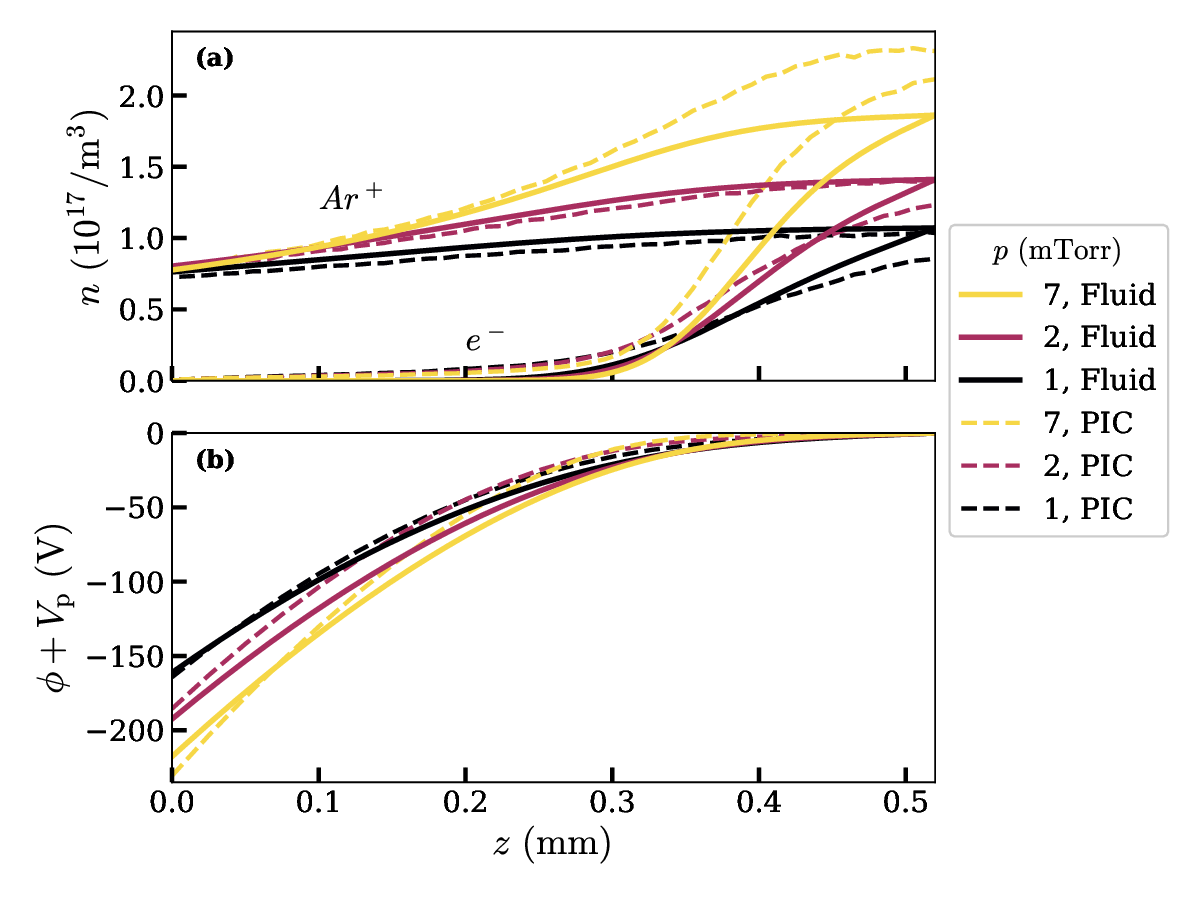}}
\caption{\label{fig:fluSheath} Solutions of the 1D fluid sheath model for 1, 2, and 7 mTorr (solid lines), and the corresponding PIC lineouts at $r=9$ mm (dashed lines).}
\end{figure}

We integrate Eq. (\ref{eq:sheath}) numerically, with all inputs provided by the presheath model, namely $n_s$, $u_{{\rm i},s}$, $\varepsilon_{{\rm e},s}$, $V_{\rm p}$, and the axial electric field $E_{s}$.
The results of the sheath model are overlaid on PIC lineouts in Fig.~\ref{fig:fluSheath}.
The model accurately predicts the potentials and densities for the simulated pressures.
The error in $n_{\rm i}$ near the sheath edge at high pressures is simply carryover from the presheath model.
The pressure dependence of $V_{\rm s}$ given by the fluid model is shown in Fig.~\ref{fig:vsvp}. The fluid model predicts $V_{\rm s}$ to within 10\% of the PIC result, capturing the slight increase with pressure.
We note that our model does not include pressure dependent boundary conditions at the cathode to account for recapture, which is consistent with our conclusion from Sec. \ref{sec:refl} that the pressure dependence of the presheath is more important to the sheath behavior than recapture.

We add the sheath voltage to the presheath voltage to obtain the full $V$-$p$ dependence from the fluid model.
This is shown in Fig.~\ref{fig:vp} along with our PIC result and the experimental results of \citet{depla2006discharge}.
The fluid model captures the monotonic decline and convexity that is consistently observed in experiment and agrees with the PIC results within 15\%.
We note that the fluid model requires the axial bounds of the presheath as input and a reasonable value for the axial drifts at the presheath-bulk interface.
These inputs were not, however, the source of the pressure dependence, since we used the same values for all pressures.
We expect the fluid model to be most applicable when the electron current at the cathode is small compared to the ion current and in the regime where the cyclotron frequency is much greater than the collision frequency. 

\section{Conclusion}\label{sec:conc}
In this work, we reproduced the consistently-seen experimental voltage-pressure dependence of DCMS with 2D-RZ PIC simulations.
These simulations showed that as pressure increased the presheath voltage decreased dramatically and the sheath voltage increased slightly, resulting in a net monotonic decline in the total discharge voltage.
We also ran a series of simulations in which we varied the electron reflection coefficient at the cathode while keeping the pressure constant to disentangle the effects of recapture and pressure.
These simulations showed that as reflection increased, recapture decreased and the sheath width and sheath voltage decreased.
The presheath, however, was unaffected by these changes that occurred in the sheath.
Taken together, the pressure sweep and reflection sweep showed that the $V$-$p$ dependence was dominated by the presheath, which was unaffected by variation in electron reflection and recapture at the cathode.
Until now, the leading explanation for voltage decreasing with increasing pressure has been based on the effect of recapture.
Our PIC results indicate that recapture does not account for the $V$-$p$ dependence in this regime.
Moreover, the simulations show that the discharge is sustained primarily by electron energization in the presheath.
The independence of the presheath and bulk plasma to the electron reflection coefficient is a new finding that will improve the comparison of simulation and experimental results.

To provide an explanation for the $V$-$p$ dependence we turned to a 1D axial fluid model of the presheath and sheath regions.
The presheath model assumed classical electron transport in the form of Pedersen conductivity.
Integrating this model from the presheath-bulk interface to the sheath edge yielded fluid parameters that agreed well with PIC, and most importantly captured the dependence of the presheath voltage on pressure.
We boundary matched the presheath model to a sheath model that assumed the electron density was given by the Boltzmann relation, and integrated from the sheath edge to the cathode surface.
Again, the model captured the pressure dependence of the particle densities and potential.
The fluid model produced a $V$-$p$ curve that agreed to within 15\% of experiment and our PIC results.
The fluid model shows that the presheath voltage decreases with increasing pressure to maintain the total ionization rate in the presheath, which is essentially fixed by the constant discharge current.
The fluid model also demonstrated that classical Pedersen transport accurately captures the axial electron drifts in this regime.

The $V$-$p$ dependence is one of the most consistent experimental results for DCMS. 
The efficacy of our PIC and fluid models in capturing the $V$-$p$ dependence suggests that they may be useful in explaining other phenomena in DCMS, and possibly pulsed power configurations like high-power impulse magnetron sputtering (HiPIMS) if the current plateau approaches steady state.
The fluid model is flexible in being able to accept arbitrary electron distribution functions, magnetic field profiles, and cross-sectional areas.
Therefore, it may be descriptive for other steady-state $\bf E\times B$ discharges.

\begin{acknowledgments}
This work was supported by NSF Grant No. PHY2206904. This research used resources of the National Energy Research Scientific Computing Center (NERSC), a Department of Energy User Facility using NERSC award FES-ERCAP0032427.
\end{acknowledgments}

\section*{Author Declarations}

\subsection*{Conflict of Interest Statement}
G. R. Werner is a paid consultant for Silvaco Inc.

\subsection*{Author Contributions}
\noindent
{\bf Joseph G. Theis}: Conceptualization (lead); Formal Analysis (lead); Investigation (lead); Software (lead); Visualization (lead); Writing - original draft (lead).
{\bf Gregory R. Werner}: Conceptualization (equal);  Formal Analysis (equal); Funding Acquisition (equal); Supervision (lead); Writing – review and editing (lead).
{\bf Thomas G. Jenkins}: Conceptualization (supporting); Writing – review and editing (supporting).
{\bf Daniel Main}: Conceptualization (supporting); Writing – review and editing (supporting).
{\bf John R. Cary}: Conceptualization (equal); Funding Acquisition (equal); Supervision (equal); Writing – review and editing (supporting).

\section*{Data Availability Statement}
The data that support the findings of this study are available from the corresponding author upon reasonable request.

\bibliography{magVP}
\end{document}